\documentclass[aps,twocolumn, floats,preprintnumbers,showpacs,nofootinbib,prl]{revtex4}
\usepackage{graphicx}
\usepackage{url}
\usepackage{color}
\usepackage{amsmath}
\usepackage{amsfonts}
\usepackage{amssymb}
\def\beq{\begin{equation}}
\def\eeq{\end{equation}}
\def\beqa{\begin{eqnarray}}
\def\eeqa{\end{eqnarray}}

\def\ltap{\ \raise.3ex\hbox{$<$\kern-.75em\lower1ex\hbox{$\sim$}}\ }
\def\gtap{\ \raise.3ex\hbox{$>$\kern-.75em\lower1ex\hbox{$\sim$}}\ }

\begin{document}

\title{The 2HDM-X and Large Hadron Collider Data}

\author{Yang Bai$^{a,b}$, Vernon Barger$^{a}$, Lisa L.~Everett,$^{a}$ and Gabe Shaughnessy$^{a}$
\\
\vspace{2mm}
${}^{a}$Department of Physics, University of Wisconsin, Madison, WI 53706, USA \\
${}^{b}$SLAC National Accelerator Laboratory, 2575 Sand Hill Road, Menlo Park, CA 94025, USA 
}


\begin{abstract}
We study the consistency of two Higgs doublet models  in light of the new bosonic particle discovery at the LHC. We work within a general setup that we call the 2HDM-X,  in which the quarks couple to both scalar doublets with aligned couplings such that flavor-changing neutral currents are absent at tree level.  The 2HDM-X encompasses the traditional Type I, Type II, lepton specific, and flipped models, but also provides for more general possibilities.  The best fit to the current data with a suppressed $\tau\tau$ signal and a $b\bar b$ signal of Standard Model strength is given by the 2HDM-X with specific parameter choices;  however, very good fits are also obtained within the lepton-specific model and a democratic model, the 2HDM-D, in which both the up-type and down-type quarks couple to each doublet with equal strengths.  The approach provides a general framework  in which to interpret future LHC Higgs data within extensions of the Standard Model with two Higgs doublets.
\end{abstract}
\maketitle

\noindent
{\it{\textbf{Introduction.}}}
The recent discovery of a Higgs-like particle with a mass of order 125 GeV at the Large Hadron Collider (LHC) may be the most significant crowning achievement in particle physics in several decades. If this particle is indeed a Higgs boson, it is of paramount importance in confirming spontaneous electroweak symmetry breaking  as the origin of mass. The properties of the Higgs boson at colliders are also sensitive to physics beyond the Standard Model (SM). Knowing or excluding additional new particles that affect the Higgs properties can increase our understanding of the nature of electroweak symmetry breaking. One of the best-motivated extensions of the SM is to add an additional Higgs doublet, resulting in two Higgs doublet models (2HDM's) (see~\cite{Branco:2011iw} for a recent review).   Specific realizations of 2HDM's have already been studied in light of the Higgs-like particle discovery~\cite{Craig:2012vn,Alves:2012ez,Chang:2012ve}.

There are several well-known 2HDM's that correspond to different ways of coupling the Higgs doublets to the SM fermions.   The models can be classified into two general categories depending on whether they result in Higgs-mediated flavor-changing neutral currents (FCNC) at tree level.  Within the first class of models, tree-level FCNC are forbidden by global symmetries such as Peccei-Quinn symmetry~\cite{Peccei:1977hh} or discrete symmetries.  This class of models include the Type I (2HDM-I), Type II (2HDM-II), lepton-specific (2HDM-L) and flipped 2HDM, which can be distinguished based on Yukawa coupling measurements~\cite{Barger:2009me}.  The second class of models, which are commonly known as Type III models, include Yukawa couplings of the up-type and down-type quarks to both Higgs doublets. This class of models generically have tree-level FCNC and are highly constrained by flavor observables which are consistent with the SM.

Motivated by the experimental hints of deviations of the Higgs boson properties from the SM predictions~\cite{Bernardi:2012gb} and the agreement of the flavor observables with the SM, we explore a simple 2HDM model framework that captures the dominant effects of Type III models in modifying the Higgs properties and at the same time has no tree-level FCNC.  In this setup, which we denote as the 2HDM-X, the quarks couple to both Higgs doublets with a flavor structure that is governed by the same Yukawa matrix~\cite{Pich:2009sp}.  This coupling choice guarantees the absence of tree-level FCNC due to fine-tuning, and thus {\it a priori} FCNC may reappear at higher-loop level with appreciable strength.  However, we choose this general framework because our principal guidance is the experimental Higgs data, which points to specific features that are not yet understood.   Before proceeding, we also comment that the label 2HDM-X we use here should not be confused with previous literature which also use this title in other contexts, such as \cite{DiazCruz:2010yq}.\\

\noindent
{\it{\textbf{The 2HDM-X.}}}
In this setup,  there are two complex Higgs doublets ${\bf \Phi}_d = (\Phi^0_d, \Phi^-_d)$ with hypercharge $Y=-1$ and ${\bf \Phi}_u = (\Phi^+_u, \Phi^0_u)$ with $Y=+1$. After electroweak symmetry breaking and assuming that the tree-level Higgs potential conserves CP, we have
\beqa
\renewcommand{\arraystretch}{1.5}
{\bf \Phi}_d &=& 
\left[\begin{array}{cc}
(v_d + \phi_d^r + i \, \phi_d^i)/\sqrt{2}\,, &
\Phi_d^-
\end{array} \right]\,,  \nonumber \\
 {\bf \Phi}_u &=&
\left[\begin{array}{cc}
\Phi_u^+\,, &
(v_u + \phi_u^r + i \, \phi_u^i)/\sqrt{2}
\end{array} \right] \,,
\eeqa
in which $v_u$ and $v_d$ are the usual electroweak vacuum expectation values (VEV's), with $v^2 = v^2_u + v^2_d = (246~\mbox{GeV})^2$. The ratio of the two VEV's is defined as usual to be $\tan{\beta} \equiv v_u/v_d$. In the scalar spectrum, there are two neutral scalars $h$ and $H$, one pseudo-scalar $A$, and one charged scalar $H^\pm$. Here we mainly consider the phenomenology of the lightest CP-even neutral scalar $h$, but we will also include the effects of the charged scalar on the $h$ properties as well as the experimental constraints on $H$, $A$ and $H^\pm$. The lightest CP-even scalar $h$ is given as usual by $h=-\phi^r_d\,\sin{\alpha} + \phi^r_u \cos{\alpha}$ with the mixing angle $-\pi/2\leq \alpha < \pi/2$. In the decoupling limit with $m_h \ll m_A\sim m_H \sim m_{H^\pm}$, one has $\cos{(\beta-\alpha)}={\cal O}(v^2/m_A^2)$~\cite{Gunion:2002zf}. The tree-level couplings of $h$ to the $W$ and $Z$ are
\beqa
g_{hVV} &=& g_V\,m_V\,\sin{(\beta - \alpha)} \,,
\label{eq:vectorcoupling}
\eeqa
in which $g_V = 2m_V/v$ for $V= W$ or $Z$. The Yukawa couplings of the Higgs doublets to the quarks and the charged leptons are given in the 2HDM-X without loss of generality as follows:
\beqa
- {\cal L} &=& y_u\,\overline{u}_R \, (\cos{\gamma_u}\,{\bf \Phi}_u \,+\, \sin{\gamma_u}\,\tilde{{\bf \Phi}}_d)\,Q_L    \nonumber \\
&&\hspace{-3mm}+\,y_d\, \overline{d}_R\,(\cos{\gamma_d}\,{\bf \Phi}_d \,+\, \sin{\gamma_d}\,\tilde{{\bf \Phi}}_u) \, Q_L   \nonumber \\
 &&\hspace{-3mm}+\, y_\ell\, \overline{e}_R\,{\bf \Phi}_d \, L_L \,+\, \mbox{h.c.}\,,
\eeqa
in which $y_{u, d, \ell}$ are $3\times 3$ Yukawa matrices and $\tilde{{\bf \Phi}}_{u, d}\equiv i\sigma_2 {\bf \Phi}_{u, d}^*$, and we have neglected the neutrino interactions with the Higgs for simplicity.  In the above, we have used the freedom to redefine the two linear combinations of ${\bf \Phi}_u$ and $\tilde{{\bf \Phi}}_d$~\cite{Davidson:2005cw} to eliminate the coupling of the leptons to $\tilde{\Phi}_u$.  Hence, in this basis there are two mixing angles $\gamma_{u, d}$ for the up-type and down-type quark couplings.  

The 2HDM-X parameter space reduces in specific limits to several familiar 2HDM's, as shown in Table ~\ref{tab:match}.  It is clear that the 2HDM-X will allow us to cover a broader range in the Higgs sector than any of these familiar limits.
\begin{table}[ht!]
\renewcommand{\arraystretch}{1.5}
\centerline{
\begin{tabular}{c|ccccc|c}
\hline \hline
Type     \hspace{2mm}         &     \hspace{1mm}   I          \hspace{2mm}      &  II    \hspace{2mm}  & L     \hspace{2mm}  &  Flipped   \hspace{2mm} &  Democratic   \hspace{2mm}   &     \hspace{1mm}   Fit   \hspace{2mm}    \\ \hline
$\gamma_u$  \hspace{2mm}  &  \hspace{1mm}  $\frac{\pi}{2}$  \hspace{2mm}  & 0  \hspace{2mm}   &  0  \hspace{2mm}   &   $\frac{\pi}{2}$  \hspace{2mm}  &   $\frac{\pi}{4}$  \hspace{2mm}    &      \hspace{1mm}   0.72  \hspace{2mm}   \\ \hline 
$\gamma_d$  \hspace{2mm} & \hspace{1mm}     0          \hspace{2mm}  &  0  \hspace{2mm}  & $\frac{\pi}{2}$  \hspace{2mm}   &  $\frac{\pi}{2}$  \hspace{2mm} &   $\frac{\pi}{4}$  \hspace{2mm}   &     \hspace{1mm}   1.40  \hspace{2mm}   \\ \hline 
 \hline
\end{tabular}
}
\caption{The values of mixing angles $\gamma_{u, d}$ for different types of 2HDM's. ``L" means the lepton-specific 2HDM (2HDM-L).  The ``flipped" 2HDM (2HDM-F) is similar to the 2HDM-II except that the charged leptons couple to the same Higgs as the up-type quarks.  The ``democratic'' (2HDM-D) refers to the case in which both doublets couple equally to up and down quarks. The last column contains the best fit values of the 2HDM-X in light of the current LHC data.
\label{tab:match}}
\end{table}

The lightest CP-even neutral scalar couplings to fermions, in units of the SM coupling, take the form
\beqa
&& ht\bar{t}:  \quad \frac{\cos{(\alpha + \gamma_u)} }{ \sin{(\beta + \gamma_u)} } \,,
\qquad hb\bar{b}: \quad -\frac{\sin{(\alpha - \gamma_d)} }{ \cos{(\beta - \gamma_d)} } 
\,, \;\; \\
&&  h \tau \bar{\tau}: \quad -\frac{\sin{\alpha} } {\cos{\beta} } \,,
\label{eq:fermioncoupling}
\eeqa
in which we have shown only the dominant tree-level couplings to the third-generation fermions. For the charged Higgs $H^\pm$, the Yukawa couplings are given by
\beqa
g_{H^- t \bar b} &=& \frac{\sqrt{2}}{v} \left[ m_t  \cot{(\beta + \gamma_u)}\,P_R + m_b  \tan{(\beta - \gamma_d)} \,P_L \right]  \,, \nonumber \\
g_{H^- \tau^+ \nu} &=& \frac{\sqrt{2}}{v} \left[ m_\tau \tan{\beta}\,P_L\right] \,.
\eeqa
Finally, the scalar potential is key for determining the $hH^+H^-$ coupling that arises in the $h\to \gamma\gamma$ decay.  The scalar potential parameters can be traded for the masses and angles $M_h, M_H, M_A, M_{H^\pm}$ and $\alpha$, such that the charged Higgs coupling can be written as
\beqa
g_{hH^+H^-} &=&2 {\cos(\alpha+\beta)\over  \sin 2 \beta}{M_{h}^2-M_{12}^2\over v} \nonumber \\&+&
\sin(\beta-\alpha){2M_{H^\pm}^2-M_h^2\over v},
\eeqa
where $M_{12}^2={2m_{12}^2/ \sin 2\beta}$ is the common mass scale of the heavy Higgs sector, and $m_{12}^2$ is the mass-term of the $\Phi_1^\dagger \Phi_2 + {\rm h.c.}$ term of the Lagrangian.  We note here that our derivation of this coupling has the opposite sign as compared to that of \cite{Gunion:2002zf}.

From Eq.~(\ref{eq:vectorcoupling}) and Eq.~(\ref{eq:fermioncoupling}), we see that the SM Higgs boson couplings can be recovered by choosing $\beta=\alpha+\pi/2$ and $\gamma_u =\gamma_d =0$. Generically, the couplings to weak gauge bosons are smaller in the 2HDM-X than the SM values. However, the couplings to fermions can be either larger or smaller than the corresponding SM couplings. Also, unlike the discrete 2HDM's, the couplings to the $t$-quark, the $b$-quark and the $\tau$-lepton are independent of each other. For example, by choosing $\alpha=0$ and non-zero values of $\gamma_u$ and $\gamma_d$, one can even have a vanishing $h\rightarrow \tau\bar{\tau}$ partial decay width. As we will see later, it is this freedom that provides a better fit for the 2HDM-X than other 2HDM's. 

In order to compare with the experimental measurements, it is necessary to determine the branching fractions of various Higgs decay channels in this setup. First, we determine the total width of $h$, which is given by
\beqa
\Gamma^{\rm tot}_{\rm X}\approx \frac{\sin^2{(\alpha - \gamma_d)} }{ \cos^2{(\beta - \gamma_d)} }\,\Gamma^{b}_{\rm SM} \,+\, \sin^2{(\beta - \alpha)}\,\Gamma^{W}_{\rm SM} \,,
\eeqa
where for illustrative purposes we have assumed that the total width is always dominated by the $b\bar b$ and $W^+ W^-$ channels and we have neglected subdominant contributions from $\tau\tau$ and $gg$. In our numerical calculations, we will include all the subdominant channels in the total width calculation. Given the parameter space freedom, the total width can be much larger or smaller than the total width of the SM Higgs boson. The ratio of branching ratios for several important channels, $\rho_i \equiv \mbox{BF}_i(\mbox{2HDM}_{\rm X})/\mbox{BF}_i({\rm SM})$, then take the form
\beqa
\rho_g &=& \frac{\cos^2{(\alpha + \gamma_u)} }{ \sin^2{(\beta + \gamma_u)} } \frac{\Gamma^{\rm tot}_{\rm SM}}{\Gamma^{\rm tot}_{\rm X}  }\,, \;
\rho_b =\frac{\sin^2{(\alpha - \gamma_d)} }{ \cos^2{(\beta - \gamma_d)} } \frac{\Gamma^{\rm tot}_{\rm SM}}{\Gamma^{\rm tot}_{\rm X}  }\,, \\
\rho_\tau &=&\frac{\sin^2{\alpha} }{ \cos^2{\beta} } \frac{\Gamma^{\rm tot}_{\rm SM}}{\Gamma^{\rm tot}_{\rm X}  }\,,\;\quad
\rho_V = \sin^2{(\beta -\alpha)} 
\frac{\Gamma^{\rm tot}_{\rm SM}}{\Gamma^{\rm tot}_{\rm X}  }\,, \\
\rho_\gamma &=& \left[  \frac{16\,\cos{(\alpha + \gamma_u)} }{ 47\,\sin{(\beta + \gamma_u)} } - \frac{63\,\sin{(\beta -\alpha)}}{47} \right]^2
\frac{\Gamma^{\rm tot}_{\rm SM}}{\Gamma^{\rm tot}_{\rm X}  } \,.
\eeqa
For $\rho_\gamma$, we show here the result in the limit $m_h \ll 2m_t, 2M_W$, but in our numerical calculations we use the full one-loop formula including the $H^\pm$ contribution. 

The experimental data are typically presented as ratios of the production cross section times branching fraction divided by the corresponding SM value ($\mu_j(i)$ for the production type ``$i$" and the decay channel ``$j$"). For the Higgs resonance and using the Breit-Wigner formula, the $\mu_j(i)$ can be related to the $\rho_i$  by $\mu_j(i) = \rho_j \, \rho_i$ for the exclusive channel. For the inclusive channel, we sum over the $gg$, $VV$ and $Vh$ production modes.\\

\noindent
{\it{\textbf{Fit to the Higgs data.}}}
We perform a Bayesian fit to the available LHC and Tevatron data following the same method as \cite{Low:2012rj}.  We require all scalar potential parameters to be perturbative with $\lambda_i^2 < 4\pi$.  From the present experimental measurements~\cite{atlas2012,cmshig12020,Tevatron} (see also \cite{Bernardi:2012gb}), a simple combination of inclusive and selected exclusive channels provides the following distilled measurements:
\begin{eqnarray}
\mu_{pp}(\gamma\gamma)&=& 2.0^{+0.4}_{-0.5} \,, \quad
\mu_{gg}(\gamma\gamma)= 1.6^{+0.5}_{-0.5}  \,, \nonumber \\
\mu_{VV}(\gamma\gamma)&=& 2.3^{+1.2}_{-1.0} \,, \quad
\mu_{pp}(VV)= 1.0^{+0.3}_{-0.3}  \,,  \nonumber\\
\mu_{gg}(VV)&=& 0.7^{+0.5}_{-0.5} \,, \quad
\mu_{VV}(VV)= 0.3^{+1.5}_{-1.6} \,,  \nonumber\\
\mu_{pp}(\tau\tau)&=& 0.5^{+1.6}_{-2.1} \,, \quad
\mu_{gg}(\tau\tau)= 1.3^{+1.1}_{-1.1} \,,  \nonumber\\
\mu_{VV}(\tau\tau)&=& -1.8^{+1.0}_{-1.0} \,, \quad
\mu_{Vh}(b\bar b) = 1.3^{+1.8}_{-0.8} \,,
\label{eq:measurements}
\end{eqnarray}
in which 
$pp, gg, VV$ and $Vh$ indicate inclusive, gluon fusion, vector boson fusion and associated production of the Higgs boson, respectively.  Furthermore, we constrain the $H$ state with the excluded signal rate over a mass range up to 500 GeV~\cite{atlas2012}.

\begin{figure}[t]
\begin{center}
\includegraphics[scale=0.5,angle=0]{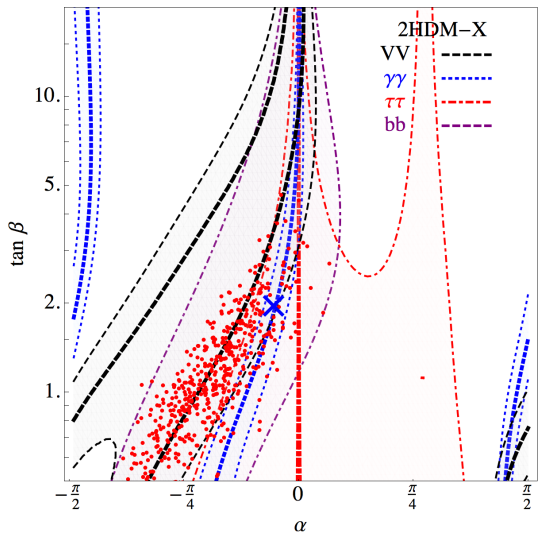}
\caption{Ranges of $\alpha$ and $\tan \beta$ consistent with the experimental data.  A random selection of points in our Bayesian fit are presented as red dots.  The point with the best overall fit is marked by the blue $\times$.  Data consisting of decays to $\gamma\gamma$, $VV$, $b\bar b$ and $\tau\tau$ constrain the parameter space to regions of black, blue, red, and purple colors, respectively.  }
\label{fig:modelfit}
\end{center}
\end{figure}

It is well known that a light charged Higgs boson can increase the $h\to \gamma\gamma$ rate.  Therefore, we include the values of the precision electroweak observables $S$ and $T$ and the flavor changing decays $B_0\to X_s+\gamma+X$ at next to leading order in our fit~\cite{Ciuchini:1997xe}.  We require $S=0.04\pm 0.09$, $T = 0.07 \pm 0.08$ with an 88\% positive correlation~\cite{Beringer:1900zz} and ${\rm BF}(B_0\to X_s+\gamma+X) = (3.55\pm0.26)\times 10^{-4}$~\cite{Amhis:2012bh}.

In addition to the enhancement in the $\gamma\gamma$ channel, a curious feature of the measurements is an apparent suppression in the $\tau\tau$ channel.  We will see that this can be accommodated in the traditional lepton-specific model as well as the democratic and general models.  

In Fig.~\ref{fig:modelfit}, we show the bands in $\alpha$ and $\tan\beta$ near the best fit point associated with the collider measurements in Eq.~(\ref{eq:measurements}).  The solid lines indicate the central value while the shaded region indicates the $1\sigma$ band of the  measurements combined over all production channels [$\mu(VV) = 0.9^{+0.2}_{-0.2}$, $\mu(\gamma\gamma) = 1.9^{+0.3}_{-0.3}$, $\mu(b\bar b) = 1.3^{+1.8}_{-0.8}$ and $\mu(\tau\tau) = 0.0 ^{+ 0.4}_{-0.0}$].  The Bayesian fit is represented by the red points.  Therefore, the measured $VV$, $\gamma\gamma$ and $\tau\tau$ rates provide a good fit for large $\tan\beta$ and $\alpha \to 0$.

\begin{figure}[t]
\begin{center}
\includegraphics[scale=0.5,angle=0]{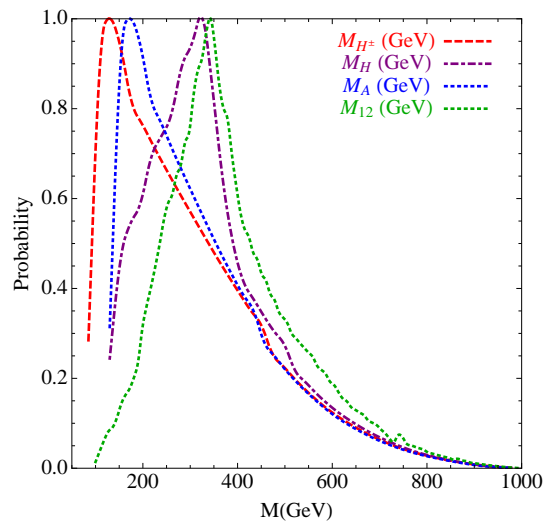}
\caption{Posterior probability of the heavy Higgs masses.  To achieve the enhanced $h\to \gamma\gamma$ rate, the charged Higgs is typically light.  Consistency with electroweak precision observables forces the CP-odd to be of similar mass. }
\label{fig:Mheavy}
\end{center}
\end{figure}

The best fit point in $\alpha$ and $\tan \beta$ is represented by the blue $\times$ and has a great overall fit of $\chi^2/\nu \approx 0.5$.  Associated with the best fit, we find a small $M_{H^\pm}$ that satisfies the current direct search bound~\cite{Searches:2001ac}, driven by the need to fit the large $h\to \gamma\gamma$ rate.  In Fig.~\ref{fig:Mheavy}, we show the posterior mass distribution of the heavy states, which have the peak values different from the decoupling limit.  Note that it is difficult to get such a large $b\bar b$ value of $\mu_{Vh}(b\bar b) = 1.3$ indicated by the data.  This is due to the already large branching fraction in the SM.  Hence, no central value line is shown for that channel.  However, this large value is dominated by the Tevatron $Vh\to b\bar b$ measurement.  If the future rate measured at the LHC is smaller as indicated by the CMS measurement, agreement within this channel for the 2HDM-X will improve.  With a suppressed $b\bar b$ rate, another means to enhance the $h\to\gamma\gamma$ rate is available as the total width decreases, thereby increasing the branching fraction to the other decay modes.  However, we find this occurs within the 2HDM-X infrequently.

\begin{figure}[t]
\begin{center}
\includegraphics[scale=0.5,angle=0]{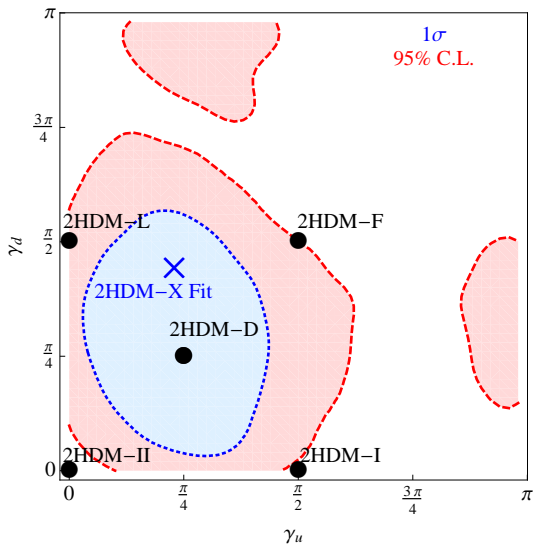}
\caption{Posterior probability regions of $1\sigma$ and 95\% C.L. in the plane of $\gamma_u$ and $\gamma_d$.  Traditional 2HDM models (black circles) the best fit of 2HDM-X (blue $\times$) and 2HDM-D (middle black circle) are shown for comparison. }
\label{fig:gamugamd}
\end{center}
\end{figure}

In Fig.~\ref{fig:gamugamd}, we show the posterior probability density in the plane of $\gamma_u$ and $\gamma_d$.  The most preferred values do not fit any of the traditional models.  Of these models, the 2HDM-II has the most tension as it lies outside the 95\% C.L. region.  The lepton-specific (2HDM-L) case provides the best fit among the traditional models due to the independence of the $\tau\tau$ and $q\bar q$ couplings.  This is verified by the reduced $\chi^2$ distribution by channel tabulated in Table~\ref{table:chi2}.  We note that the values of $\gamma_u=\gamma_d=\pi/4$ appear to give quite good fits for the democratic 2HDM-D model.  Among all the models we consider, the $\tau\tau$ channel provides the most tension.  However, the 2HDM-L,  2HDM-D, and the best-fit point of the 2HDM-X model provide excellent fits to the lepton data.

\begin{figure}[t]
\begin{center}
\includegraphics[scale=0.5,angle=0]{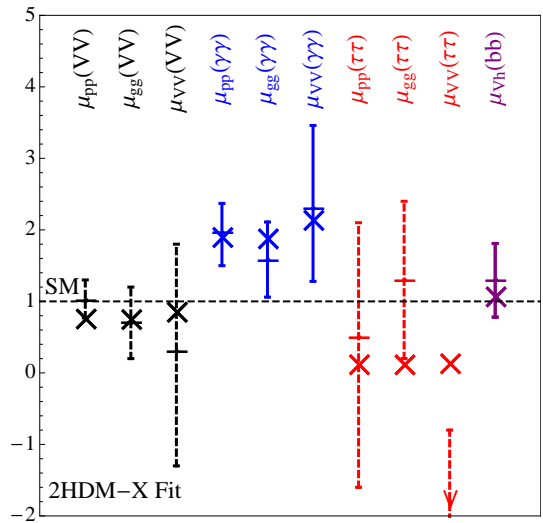}
\caption{Channels measured at the LHC and Tevatron with best fit points from the 2HDM-X model denoted by $\times$.}
\label{fig:pull}
\end{center}
\end{figure}
\begin{table}[h]
\caption{Reduced $\chi^2$ values for key channels in our analysis.  For each decay channel, we sum over the production mechanisms.  }
\begin{center}
\begin{tabular}{|c|cccc|c|}
\hline
Model & $\chi^2_{VV}/\nu$ & $\chi^2_{\gamma\gamma}/\nu$& $\chi^2_{b\bar b}/\nu$ & $\chi^2_{\tau\tau}/\nu$ &$\chi^2_{\rm total}/\nu$\\
\hline
\hline
SM & 0.39& 0.96 & 0.71 & 2.29 & 1.01\\
Type-I & 0.54& 0.48 & 1.37 & 2.48 & 0.75\\
Type-II & 0.18& 0.23 & 0.59 & 2.14 & 0.68\\
Lepton & 0.47& 0.30 & 0.68 & 1.32 & 0.59\\
Flipped & 0.19& 0.19 & 0.59 & 1.87 & 0.61\\
\hline
\hline
Democratic & 0.20& 0.23 & 0.83 & 1.31 & 0.53\\
General & 0.18& 0.20 & 0.60 & 1.31 & 0.50\\
\hline
\end{tabular}
\end{center}
\label{table:chi2}
\end{table}%
As shown in Table~\ref{table:chi2}, the best fit over all the data is given by the 2HDM-X, with the 2HDM-D not far behind.  The pull from the data of the best fit point in the general model is indicated in Fig.~\ref{fig:pull}.  We immediately see that the $\tau\tau$ gives negligible rate while the $\gamma\gamma$ rate is well matched.  It is worth mentioning that while the SM still gives an acceptable $\chi^2/\nu\approx 1$, the large tension within the $\tau\tau$ and $\gamma\gamma$ channels that persists even for the more traditional 2HDM warrant consideration of the more general model as a potential explanation of the data.

\noindent
{\it{\textbf{Conclusions.}}}
The 2HDM-X framework provides a rich setting in which to analyze the LHC Higgs-like data in the context of simple extensions of the SM  to allow for two electroweak Higgs doublets.  As one might expect,  the general 2HDM-X provides the best fit to the current data, with the democratic and lepton-specific models not far behind. A light charged Higgs is indicated at the best fit point in the 2HDM-X.  Further experimental measurements of the neutral Higgs boson properties and searches for the charged Higgs boson will verify or falsify models within the 2HDM-X framework.  

We note here that during the completion of our paper, another paper appeared that also analyzes general two Higgs doublet models with vanishing tree-level FCNC's, with a different parametrization~\cite{Gori}.

{\it{\textbf{Acknowledgements.}}}
We thank Daniel Chung  for useful discussions. VB, LE, and GS are supported by  the U.~S.~Department of Energy under the contract DE-FG-02-95ER40896. YB is supported by start-up funds from the Univ. of Wisconsin, Madison. YB thanks SLAC for their warm hospitality.


\begin{thebibliography}{99}
\bibitem{Branco:2011iw} 
  G.~C.~Branco, P.~M.~Ferreira, L.~Lavoura, M.~N.~Rebelo, M.~Sher and J.~P.~Silva,
  Phys.\ Rept.\  {\bf 516}, 1 (2012)
  [arXiv:1106.0034 [hep-ph]].
  
\bibitem{Craig:2012vn} 
  N.~Craig and S.~Thomas,
  arXiv:1207.4835 [hep-ph].


\bibitem{Alves:2012ez} 
  D.~S.~M.~Alves, P.~J.~Fox and N.~J.~Weiner,
  arXiv:1207.5499 [hep-ph].
  
\bibitem{Chang:2012ve} 
  S.~Chang, S.~K.~Kang, J.~-P.~Lee, K.~Y.~Lee, S.~C.~Park and J.~Song,
  arXiv:1210.3439 [hep-ph].
  
\bibitem{Peccei:1977hh} 
  R.~D.~Peccei and H.~R.~Quinn,
  Phys.\ Rev.\ Lett.\  {\bf 38}, 1440 (1977).
  
  
\bibitem{Barger:2009me} 
  V.~Barger, H.~E.~Logan and G.~Shaughnessy,
  Phys.\ Rev.\ D {\bf 79}, 115018 (2009)
  [arXiv:0902.0170 [hep-ph]].
  
  
\bibitem{Bernardi:2012gb} 
  G.~Bernardi and M.~Herndon,
  arXiv:1210.0021 [hep-ex].
  

 




\bibitem{Pich:2009sp} 
  A.~Pich and P.~Tuzon,
  Phys.\ Rev.\ D {\bf 80}, 091702 (2009)
  [arXiv:0908.1554 [hep-ph]].
  
    \bibitem{DiazCruz:2010yq} 
  J.~L.~Diaz-Cruz, A.~Diaz-Furlong and J.~H.~Montes de Oca,
  arXiv:1010.0950 [hep-ph].
  
\bibitem{Gunion:2002zf} 
  J.~F.~Gunion and H.~E.~Haber,
  Phys.\ Rev.\ D {\bf 67}, 075019 (2003)
  [hep-ph/0207010].


\bibitem{Davidson:2005cw} 
  S.~Davidson and H.~E.~Haber,
  Phys.\ Rev.\ D {\bf 72}, 035004 (2005)
  [Erratum-ibid.\ D {\bf 72}, 099902 (2005)]
  [hep-ph/0504050].



\bibitem{Low:2012rj} 
  I.~Low, J.~Lykken and G.~Shaughnessy,
  arXiv:1207.1093 [hep-ph].
   
\bibitem{atlas2012} 
  G.~Aad {\it et al.}  [ATLAS Collaboration],
  arXiv:1207.0319 [hep-ex];
  arXiv:1207.7214 [hep-ex].

\bibitem{cmshig12020} 
[CMS Collaboration], CMS PAG HIG-12-020;
CMS PAG HIG-12-015;
 CMS PAG HIG-12-018.

\bibitem{Tevatron}
  [Tevatron New Physics Higgs Working Group and CDF and D0 Collaborations],
  arXiv:1207.0449 [hep-ex].
  

\bibitem{Ciuchini:1997xe} 
  M.~Ciuchini, G.~Degrassi, P.~Gambino and G.~F.~Giudice,
  Nucl.\ Phys.\ B {\bf 527}, 21 (1998)
  [hep-ph/9710335].
  
\bibitem{Beringer:1900zz} 
  J.~Beringer {\it et al.}  [Particle Data Group Collaboration],
  Phys.\ Rev.\ D {\bf 86}, 010001 (2012).
  
\bibitem{Amhis:2012bh} 
  Y.~Amhis {\it et al.}  [Heavy Flavor Averaging Group Collaboration],
  arXiv:1207.1158 [hep-ex].
  
\bibitem{Searches:2001ac} 
  [LEP Higgs Working Group for Higgs boson searches and ALEPH and DELPHI and L3 and OPAL Collaborations],
  hep-ex/0107031.

\bibitem{Gori} 
  W.~Altmannshofer, S.~Gori and G.~D.~Kribs,
  arXiv:1210.2465 [hep-ph].
 





\end{thebibliography}
\end{document}